\documentclass[doublecol]{epl2}

\usepackage{latexsym}
\usepackage{amsfonts}
\usepackage{amssymb}
\usepackage{amsmath}
\usepackage{graphicx}

\title{Scanning tunneling spectroscopic evidence for magnetic field-induced
microscopic orders in the high-$T_c$ superconductor YBa$_2$Cu$_3$O$_{7-\delta}$}
\shorttitle{Magnetic field-induced microscopic orders in YBa$_2$Cu$_3$O$_{7-\delta}$}

\author{A.D. Beyer\inst{1} \and M.S. Grinolds\inst{1} \and M.L. Teague\inst{1} \and S. Tajima\inst{2} \and N.-C. YEH\inst{1}}
\shortauthor{A.~D. Beyer \etal}

\institute{
  \inst{1} Department of Physics, California Institute of Technology, Pasadena, CA 91125, USA\\
  \inst{2} Department of Physics, Osaka University, Osaka 560-0043, Japan}

\abstract{We report spatially resolved tunneling spectroscopic evidence for field-induced microscopic orders in a high-$T_c$ superconductor $\rm YBa_2Cu_3O_{7-\delta}$. The spectral characteristics inside vortices reveal a pseudogap ($V_{\rm CO}$) larger than the superconducting gap ($\Delta _{\rm SC}$) as well as a subgap ($\Delta ^{\prime}$) smaller than $\Delta _{\rm SC}$, and the spectral weight shifts steadily from $\Delta _{\rm SC}$ to $V_{\rm CO}$ and $\Delta ^{\prime}$ upon increasing magnetic field. Additionally, energy-independent conductance modulations at 3.6 and 7.1 lattice constants along the Cu-O bonding directions and at 9.5 lattice constants along the nodal directions are manifested in the vortex state. These wave-vectors differ fundamentally from the strongly dispersive modes due to Bogoliubov quasiparticle scattering interferences and may be associated with field-induced microscopic orders of pair-, charge- and spin-density waves.}

\begin{document}

\maketitle

\section{Introduction}
In conventional type-II superconductors, superconductivity is suppressed inside periodic Abrikosov vortices~\cite{Abrikosov57}, leading to continuous quasiparticle bound states and a peak of local density of states (LDOS) at zero energy~\cite{Caroli64,Gygi91,Hess90}. In contrast, the effect of magnetic field on high-$T_c$ superconductors is much more complicated than that on conventional type-II superconductors. Macroscopically, high-$T_c$ cuprates are extreme type-II superconductors with strong thermal, disorder and quantum fluctuations~\cite{FisherDS91,Blatter94,Yeh05,Beyer07}. Microscopically, neutron scattering experiments on hole-doped cuprate $\rm La_{1.84}Sr_{0.16}CuO_4$ reported an effective radius of vortices substantially larger than the superconducting coherence length $\xi _{\rm SC}$~\cite{Lake01}. Scanning tunneling spectroscopic (STS) studies of $\rm Bi_2Sr_2CaCu_2O_{8+x}$ (Bi-2212) found no zero-bias conductance peaks inside vortices~\cite{Renner98,Pan00}. Further detailed spatially resolved STS studies of Bi-2212 in one magnetic field $H = 5$ T revealed a field-induced $(4a_0 \times 4a_0)$ conductance modulation inside each vortex, where $a_0$ = 0.385 nm is the planar lattice constant of Bi-2212~\cite{Hoffman02}. The latter finding has been attributed to the presence of a coexisting competing order (CO) such as pair-density waves (PDW)~\cite{ChenHD02,ChenHD04}, pinned spin-density waves (SDW)~\cite{Demler01,Polkovnikov02}, or charge-density waves (CDW)~\cite{Kivelson03,LiJX06,ChenCT07,Beyer08,Boyer07} upon suppression of SC inside the vortices. However, there have not been high-resolution STS studies of the field-induced collective modes on other high-$T_c$ superconductors than Bi-2212, neither has there been detailed investigation of $H$-dependent evolution of the collective modes, although vortex images had been reported in Y-123~\cite{Maggio-Aprile95,Fischer07} and effects of quasiparticle scattering by vortices have also been investigated in $\rm Ca_{2-x}Na_xCaO_2Cl_2$~\cite{Hanaguri09}. 

In this letter we report spatially resolved vortex-state STS studies of Y-123 as functions of magnetic field, which reveal two field-induced energy scales and three density-wave modes with energy-independent wave-vectors, in addition to the effects associated with quasiparticle scattering interferences~\cite{McElroy05,ChenCT03}. These field-induced microscopic orders differ fundamentally from the predictions due to simple Bogoliubov quasiparticle scattering~\cite{McElroy05,ChenCT03} and are suggestive of significant interplay between SC and competing orders upon increasing magnetic field.

\section{Experimental}
The synthesis and characterization of the Y-123 untwinned single crystal with $T_c$ = 93 K studied here have been described elsewhere~\cite{Limonov00}. We use chemical etching to prepare Y-123 surface for STS experiments because Y-123 is more difficult to handle than Bi-2212~\cite{Fischer07,Vasquez94}, and bromine chemical etching techniques can reproducibly remove non-stoichiometric surface layers to reveal quality surfaces, as manifested by x-ray photoemission spectroscopy~\cite{Vasquez94}. 

The spatially resolved tunnelling conductance ($dI/dV$) versus energy ($\omega = eV$) spectra for the quasiparticle LDOS maps were obtained with our homemade cryogenic scanning tunnelling microscope (STM). Our STM has a base temperature of 6 K, variable temperature range up to room temperature, magnetic field range up to 7 Tesla, and ultra-high vacuum capability down to a base pressure $< 10^{-9}$ Torr at 6 K. For each constant temperature ($T$) and magnetic field ($H$), the experiments were conducted by tunneling currents along the crystalline c-axis under a range of bias voltages at a given location. The typical junction resistance was $\sim 1$ G$\Omega$. Current ($I$) vs. voltage ($V$) measurements were repeated pixel-by-pixel over an extended area of the sample. To remove slight variations in the tunnel junction resistance from pixel to pixel, the differential conductance at each pixel is normalized to its high-energy background~\cite{Yeh01}.

\section{Results and Analysis}
In Fig.~1(a) we illustrate the normalized zero-field c-axis tunneling conductance spectra taken at $T$ = 6, 77 and 102 K. For $T$ = 6 K, the spectrum exhibits clear coherence peaks at energies $\pm \Delta_{\rm SC} \sim \pm$ 20 meV and shoulder-like satellite features at $\pm \Delta_{\rm eff} \sim \pm$ 38 meV. At $T$ = 77 K $< T_c$, only one set of rounded features remains. Eventually for $T$ = 102 K $> T_c$ the peaks vanish within experimental resolution. The long-range homogeneity of the zero-field tunneling spectra is exemplified by the $(95 \times 95)$ nm$^2$ spatial map of the superconducting (SC) gap $\Delta _{\rm SC}$ in Fig.~1(b) and by the corresponding energy histogram ($\Delta _{\rm SC} = 20.0 \pm 1.0$ meV) in Fig.~1(c), which differs from the strong spatial variations in the quasiparticle spectra of Bi-2212~\cite{McElroy05}. In contrast, the satellite features at $\pm \Delta_{\rm eff}$ exhibit stronger spatial inhomogeneity, as manifested by the $(95 \times 95)$ nm$^2$ spatial map of $\Delta_{\rm eff}$ in Fig.~1(e) and by the corresponding histogram in Fig.~1(f), showing $\Delta_{\rm eff} = 37.8 \pm 1.2$ meV.

\begin{figure}
\centering
\includegraphics[width=3.4in]{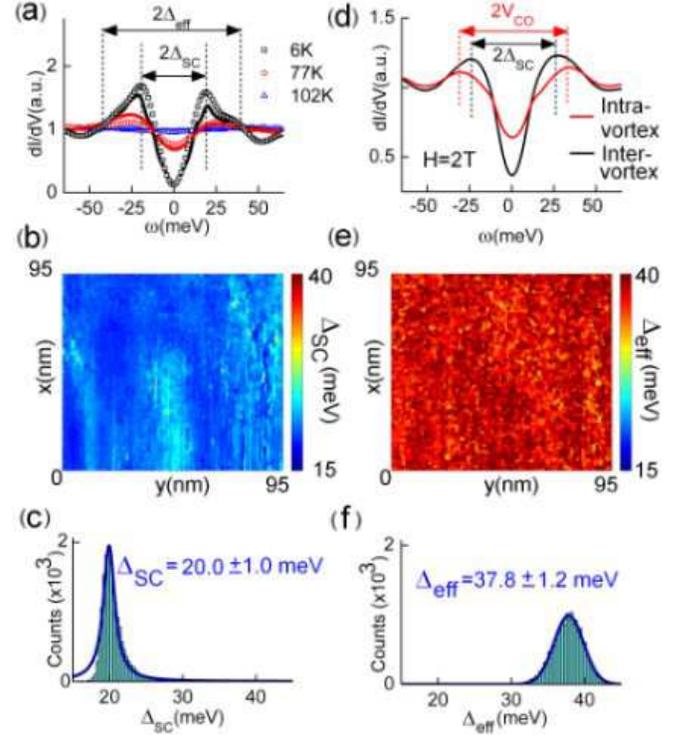}
\caption{(color online) Implication of CO from zero- and finite-field STS in Y-123: (a) Normalized zero-field tunneling spectra taken at $T$ = 6, 77 and 102 K. The solid lines represent fittings to the $T$ = 6, 77 and 102 K spectra by assuming coexisting SC and CO, following Refs.~\cite{ChenCT07,Beyer08}. For more details of the data normalization and theoretical fitting procedures, see Discussion and Refs.~\cite{ChenCT07,Beyer08}. (b) The $\Delta _{\rm SC}$ map over a $(95 \times 95)$ nm$^2$ area at $T$ = 6 K and $H$ = 0. (c) Histogram of $\Delta _{\rm SC}$ over the same area as in (b), showing $\Delta _{\rm SC} = (20.0 \pm 1.0)$ meV. (d) Spatially averaged intra- and inter-vortex spectra for $T$ = 6 K and $H$ = 2 T. (e) The $\Delta_{\rm eff}$ map over a $(95 \times 95)$ nm$^2$ area at $T$ = 6 K and $H$ = 0. (f) Histogram of $\Delta_{\rm eff}$ in the same area as in (e), showing $\Delta_{\rm eff} = 37.8 \pm 1.2$ meV.}
\label{Fig1}
\end{figure}

A possible interpretation for the zero-field spectra in Fig.~1(a) may be a scenario of coexisting CO and SC in the ground state of the cuprates~\cite{ChenCT07,Beyer08,Boyer07,ChenCT03,Vershinin04}. Following the analysis briefly outlined in the Discussion section and detailed elsewhere~\cite{ChenCT07,Beyer08}, we can account for the zero-field spectra in Fig.~1(a) by incorporating realistic bandstructures for Y-123 and assuming either CDW or disorder-pinned SDW as the CO with a wave-vector $\textbf{Q}_{\rm CO}$ parallel to the Cu-O bonding directions. We find that both CDW and disorder-pinned SDW yield equally good fitting except $|\textbf{Q}_{\rm CDW}| = 2|\textbf{Q}_{\rm SDW}|$~\cite{Polkovnikov02}. At $T \ll T_c$ our theoretical fitting (black solid curve) can account for the sharp superconducting coherence peaks at $\omega = \pm \Delta_{\rm SC}$ and the shoulder-like satellite features at $\omega = \pm \Delta_{\rm eff}$, where the effective gap $\Delta_{\rm eff}$ is related to $\Delta_{\rm SC}$ and the CO energy $V_{\rm CO}$ via the relation $\Delta _{\rm eff} ^2 = \lbrack \Delta _{\rm SC} ^2 + V_{\rm CO} ^2 \rbrack$.~\cite{ChenCT07,Beyer08} Thus, we obtain $\Delta_{\rm SC}$ = 20 meV, $V_{\rm CO}$ = 32 meV and $|\textbf{Q}_{\rm CO}| = (0.25 \pm 0.03)\pi$. For elevated temperatures below $T_c$, such as $T$ = 77 K, our fitting (red solid curve in Fig.~1(a)) also agrees with experimental data with rounded features at $\pm \Delta_{\rm eff}(T)$~\cite{ChenCT07,Beyer08}.

An alternative way to verify the feasibility of the CO scenario is to introduce vortices because the suppression of SC inside vortices may unravel the spectroscopic characteristics of the remaining CO. As exemplified in Fig.~1d for a set of intra- and inter-vortex spectra taken at $H$ = 2 T, the quasiparticle spectra near the center of each vortex exhibit pseudogap (PG)-like features, in contrast to theoretical predictions for a sharp zero-energy peak around the center of the vortex core had superconductivity been the sole order in the ground state~\cite{Caroli64,Gygi91,Franz98}. Interestingly, the PG energy inside vortices is comparable to the CO energy $V_{\rm CO} \approx 32$ meV derived from our zero-field fitting as well as the spin gap energy obtained from neutron scattering studies of optimally doped Y-123~\cite{DaiP01}.

\begin{figure}
\centering
\includegraphics[width=3.4in]{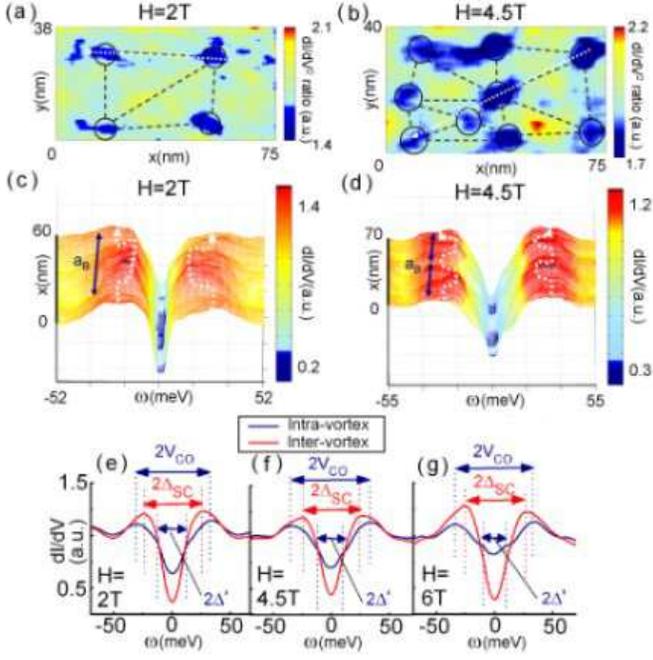}
\caption{(color online) Vortex-state conductance maps at $T$ = 6 K in Y-123: (a) Conductance power ratio $r_G$ map over a $(75 \times 38)$ nm$^2$ area for $H$ = 2 T, showing disordered vortices with an average $a_B = (33.2 \pm 9.0)$ nm. (b) The $r_G$ map over a $(75 \times 40)$ nm$^2$ area for $H$ = 4.5 T, showing $a_B = (23.5 \pm 8.0)$ nm. (c) Conductance spectra along the white line in (a), showing SC peaks at $\omega = \pm \Delta _{\rm SC}$ outside vortices and PG features at $\omega = \pm V_{\rm CO}$ inside vortices. (d) Conductance spectra along the dashed line indicated in (b). Spatially averaged intra- and inter-vortex spectra for (e) $H$ = 2 T, (f) $H$ = 4.5 T and (g) $H$ = 6 T. }
\label{Fig2}
\end{figure}

To investigate how quasiparticle spectra evolve with field, we performed spatially resolved spectroscopic studies at $H$ = 2, 4.5, 5 and 6 T and for $T$ = 6 K. In Figs.~2(a) and 2(b) we show exemplified spatial maps of the conductance power ratio $r_G$ for $H$ = 2 and 4.5 T, respectively. Here $r_G$ at every pixel is defined as the ratio of the conductance power $(dI/dV)^2$ at $\omega  = \Delta _{\rm SC}$ relative to that at $\omega = 0$. We find that the presence of vortices is associated with the local minimum of $r_G$ because of enhanced zero-energy quasiparticle density of states inside the vortex core. Moreover, the total flux is conserved within the area studied despite the appearance of disordered vortices. That is, the total number of vortices multiplied by the flux quantum is equal to the magnetic induction multiplied by the area, within experimental errors. Thus, we obtain averaged vortex lattice constants $a_B = 33.2$ nm and 23.5 nm for $H$ = 2 T and 4.5 T, respectively, comparable to the theoretical values of $a_B = 35.0$ nm and 23.3 nm. On the other hand, the mean ``vortex halo'' radius $\xi _{\rm halo}$ is much longer than the SC coherence length $\xi _{\rm SC}$, and the average $\xi _{\rm halo}$ decreases with field. We find $\xi _{\rm halo} = (7.7 \pm 0.3)$ nm for $H$ = 2 T, $(6.4 \pm 0.6)$ nm for $H$ = 4.5 T, and $(5.0 \pm  0.7)$ nm for $H$ = 6 T.

\begin{figure*}
\begin{center}
\includegraphics[width=6.4in]{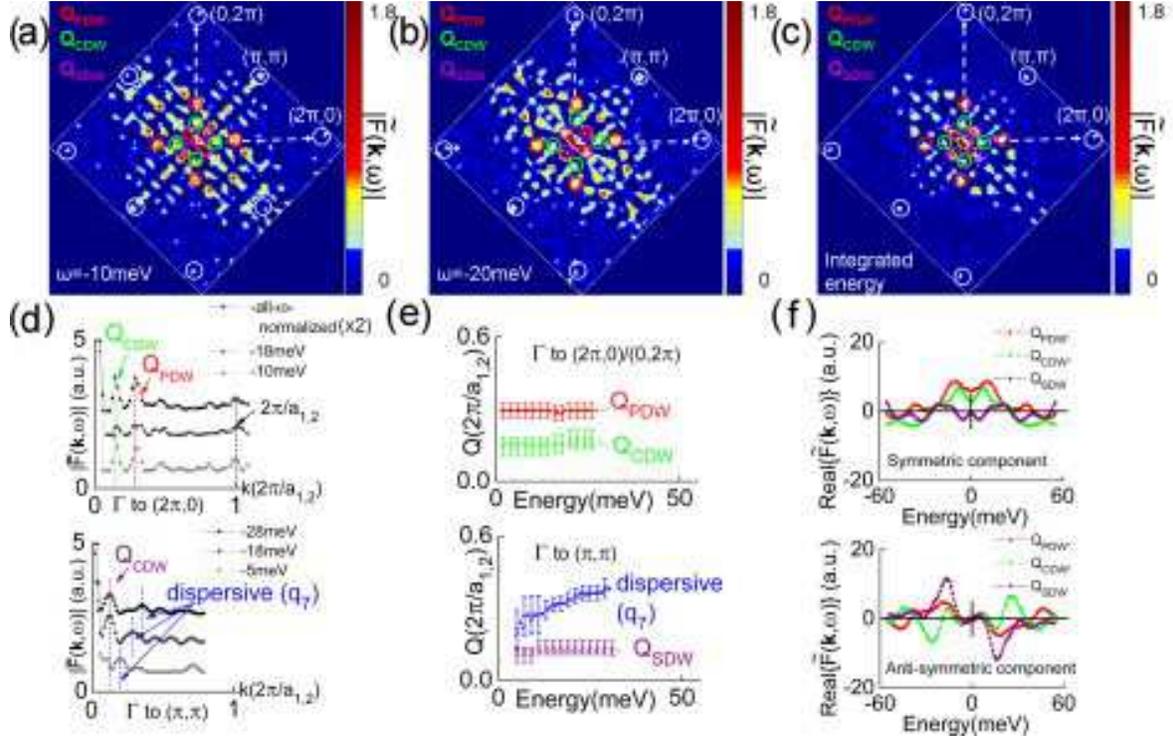}
\caption{(color online) FT studies of the vortex-state conductance maps in the two-dimensional reciprocal space and at $H$ = 5T: (a) FT-LDOS $|\tilde{F} (\textbf{k}, \omega)|$ for $\omega = -10$ meV. (b) FT-LDOS $|\tilde{F} (\textbf{k}, \omega)|$ for $\omega = -20$ meV. (c) Normalized FT-LDOS obtained by integrating $|\tilde{F} (\textbf{k}, \omega)|$ from $-1$ meV to $-30$ meV. Comparing (a) -- (c), we find three sets of energy-independent spots in addition to the reciprocal lattice constants and the $(\pi , \pi)$ resonance: $\textbf{Q}_{\rm PDW}$ and $\textbf{Q}_{\rm CDW}$ along the $(\pi,0)/(0,\pi)$ directions and $\textbf{Q}_{\rm SDW}$ along $(\pi,\pi)$, which are circled for clarity. (d) $|\tilde{F} (\textbf{k}, \omega)|$ for different energies are plotted against $\textbf{k} \parallel (\pi,0)$ and $(\pi,\pi)$ in the upper and lower panels, respectively, showing peaks at energy-independent $\textbf{Q}_{\rm PDW}$, $\textbf{Q}_{\rm CDW}$ and the reciprocal lattice constants at $(2 \pi/a_1)$ along $(\pi,0)$, and at $\textbf{Q}_{\rm SDW}$ along $(\pi, \pi)$. Additionally, dispersive wavevectors due to quasiparticle scattering interferences are found, as exemplified in the lower panel. (e) Momentum ($|\textbf{q}|$) vs. energy ($\omega$) for $|\textbf{Q}_{\rm PDW}|$, $|\textbf{Q}_{\rm CDW}|$ and $|\textbf{Q}_{\rm SDW}|$. One dispersive wavevector along $(\pi,\pi)$, denoted as $q_7$~\cite{McElroy05}, is also shown in the lower panel for comparison. (f) The symmetric and anti-symmetric components of Re$\lbrack \tilde{F} (\textbf{k}, \omega) \rbrack$ for $\textbf{k} = \textbf{Q}_{\rm PDW}$, $\textbf{Q}_{\rm CDW}$ and $\textbf{Q}_{\rm SDW}$ are shown as functions of $\omega$ in the upper and lower panels, respectively.}
\label{Fig3}
\end{center}
\end{figure*}

The spatial evolution of the vortex-state spectra may be better manifested by following a line through multiple vortices in the vortex maps of Figs.~2(a) and 2(b). As shown in Figs.~2(c) and 2(d) for $H$ = 2 and 4.5 T, respectively, the vortex-state spectral characteristics differ from those of conventional type-II superconductors~\cite{Hess90}, showing modulating gap-like features everywhere without any zero-energy peaks. In Figs.~2(e) - 2(g) we compare representative spectra taken inside and outside of vortices for $H$ = 2, 4.5 and 6 T. In a constant field, the inter-vortex spectrum reveals a sharper set of peaks at $\omega = \pm \Delta _{\rm SC}$, whereas the intra-vortex spectrum exhibits PG features at $\omega = \pm V_{\rm CO}$ and $V_{\rm CO} > \Delta _{\rm SC}$. Additional subgap features at $\omega = \pm \Delta ^{\prime} = \pm (7 \sim 10)$ meV are found inside vortices, which become more pronounced with increasing $H$. The physical origin of $\Delta ^{\prime}$ is still unknown. 

Next, we perform Fourier transformation (FT) of the vortex-state spectra and compare the results with the FT zero-field spectra taken in the same area. We define the FT-LDOS taken under a magnetic field $H$ and at a constant energy $\omega$ by the quantity $F (\textbf{k}, \omega , H)$. In Figs.~3(a) -- 3(c), we illustrate field-induced FT-LDOS $|\tilde{F} (\textbf{k}, \omega , H) | \equiv |F(\textbf{k}, \omega , H) - F(\textbf{k}, \omega ,0)|$ at $H$ = 5 T and integrated over different ranges of energies, where
\begin{equation}
\tilde{F} (\textbf{k}, \omega , H) \equiv \sum _i e^{i \textbf{k} \cdot \textbf{R} _i} \left[ \frac{dI}{dV}(\textbf{R} _i, \omega , H) - \frac{dI}{dV}(\textbf{R} _i, \omega , 0) \right].
\label{eq:FT}
\end{equation}
Here $\textbf{R} _i$ denotes the coordinate of the $i$-th pixel, and the sum is taken over all pixels of each two-dimensional map. Systematic analysis of the energy dependence of the FT-LDOS~\cite{FTfootnote} reveals two types of diffraction spots. One type of spots are strongly dispersive and may be attributed to elastic quasiparticle scattering interferences as seen in the zero-field FT-LDOS~\cite{McElroy05,ChenCT03}. The other type of spots are nearly energy-independent, as manifested in Figs.~3(a) and 3(b) for FT-LDOS at two different energies $\omega = -10$ meV and $-20$ meV, respectively, and in Fig.~3(c) for normalized FT-LDOS integrated from $\omega = -1$ meV and $-30$ meV. In addition to the reciprocal lattice vectors, we find two sets of nearly energy-independent wave-vectors along $(\pi,0)/(0,\pi)$ and one set along $(\pi, \pi)$: $\textbf{Q}_{\rm PDW} = \lbrack \pm (0.56 \pm 0.06)\pi/a_1,0 \rbrack$ and $\lbrack 0, \pm (0.56 \pm 0.06) \pi/a_2 \rbrack$, $\textbf{Q}_{\rm CDW} = \lbrack \pm (0.28 \pm 0.02)\pi/a_1,0 \rbrack$ and $\lbrack 0, \pm (0.28 \pm 0.02) \pi/a_2 \rbrack$, and $\textbf{Q}_{\rm SDW} = \lbrack \pm (0.15 \pm 0.01)\pi/a_1 , \pm (0.15 \pm 0.01)\pi/a_2 \rbrack$. Here $a_1 =$ 0.383 nm and $a_2 =$ 0.388 nm. For clarity, we illustrate $|\tilde{F} (\textbf{k}, \omega)|$ for different energies along $\textbf{k} \parallel (\pi,0)$ in the upper of Fig.~3(d) and along $(\pi,\pi)$ in the lower panel, where the $\omega$-independent peaks correspond to $|\textbf{Q}_{\rm XDW}|$ (X = P, C, S) and the reciprocal lattice vector $(2 \pi/a_{1,2})$. These wave-vectors are nearly energy independent, as shown in Fig.~3(e). For comparison, a dispersive wavevector due to quasiparticle scattering interferences~\cite{McElroy05,ChenCT03} along $(\pi , \pi)$, which is denoted as $q_7$ in Ref.~\cite{McElroy05}, is also shown in the lower panel of Fig.~3(e). We note that the dispersion relation for the mode $q_7$ along the nodal direction is in good agreement with both the experimental results found in Bi-2212~\cite{McElroy05} and the theoretical predictions for quasiparticle scattering interferences~\cite{McElroy05,ChenCT03}. 

Interestingly, we find that $\textbf{Q}_{\rm PDW}$ and $\textbf{Q}_{\rm CDW}$ correspond to charge modulations at wavelengths of $(3.6 \pm 0.4) a_{1,2}$ and $(7.1 \pm 0.6) a_{1,2}$ along the Cu-O bonding directions. The former is comparable to the checkerboard modulations reported in the vortex state of Bi-2212~\cite{Hoffman02}, whereas the latter is consistent with the Fermi surface-nested CDW wave-vector derived from our zero-field analysis. On the other hand, the nodal wave-vectors $\textbf{Q}_{\rm SDW}$ may be associated with SDW because of the field-induced unequal populations of spin-up and spin-down quasiparticles. In this context, we note that intense spots associated with the $(\pi , \pi)$ spin resonance~\cite{DaiP01} are also manifested for $\omega < \Delta _{\rm SC}$, as exemplified in Fig.~3(a) for $\omega = -9$ meV. These field-induced collective modes along the nodal direction are suggestive of important interplay between SC and spin excitations in the cuprates.  

To better understand the nature of these field-induced wave-vectors, we consider the symmetry of the complex quantity Re$\lbrack \tilde{F} (\textbf{k}, \omega) \rbrack$ relative to energy $(\omega)$ at $\textbf{k} = \textbf{Q}_{\rm XDW}$ in Fig.~3(f), with the symmetric and anti-symmetric components of Re$\lbrack \tilde{F} \rbrack$ at $\textbf{k} = \textbf{Q}_{\rm XDW}$ shown in the upper and lower panels, respectively. We find that $\tilde{F}$ is predominantly symmetric at $\textbf{Q} _{\rm PDW}$ and primarily antisymmetric at $\textbf{Q} _{\rm SDW}$. On the other hand, $\tilde{F}$ appears to have comparable symmetric and anti-symmetric components at $\textbf{Q} _{\rm CDW}$. The antisymmetric $\tilde{F}$ is consistent with SDW scenario for $\textbf{k} \parallel (\pi, \pi)$~\cite{Demler01,Polkovnikov02}, whereas for $\textbf{k} \parallel (\pi, 0)/(0,\pi)$, symmetric and anti-symmetric $\tilde{F}$ may be attributed respectively to PDW~\cite{ChenHD02,ChenHD04} and CDW~\cite{Kivelson03,LiJX06,ChenCT07,Beyer08,Boyer07}.

In Fig.~4(a) we illustrate the energy histograms of the SC and PG (or CO) features for $H$ = 0, 2, 4.5 and 6 T. A strong spectral shift from SC at $\omega = \Delta_{\rm SC}$ to PG at $\omega = V_{\rm CO}$ is seen with increasing $H$, together with the appearance of a third subgap (SG) feature at $\omega = \Delta ^{\prime} < \Delta_{\rm SC}$. For comparison, schematic histograms for conventional type-II superconductors are shown in Fig.~4(b), which exhibit spectral shifts from an initial $\omega = \Delta_{\rm SC}$ to a continuous energy distribution $\omega < \Delta_{\rm SC}$, with an additional peak appearing at $\omega = 0$ in the $H \ll H_{c2}$ and $T \ll T_c$ limit. The fraction of the spectral downshift is approximately given by $(\pi \xi _{\rm SC}^2/2)/(\sqrt{3}a_B ^2/4)$, which is linear in $H$. The predictions in Fig.~4(b) apparently differ from our empirical findings in Fig.~4(a). In Fig.~4(c) we summarize the Gaussian fitting parameters to the histograms in Fig.~4(a). We find that the values of both $\Delta_{\rm SC}$ and $V_{\rm CO}$ remain invariant with $H$, whereas the CO spectral weight and linewidth increase with increasing $H$.

\begin{figure}
\centering
\includegraphics[width=3.4in]{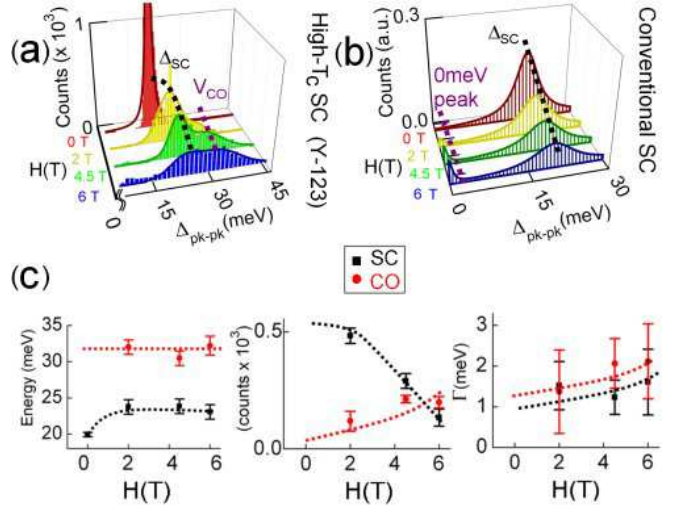}
\caption{(color online) Field-dependent spectral evolution at T = 6 K in Y-123: (a) Energy histograms derived from STS data for $H$ = 0, 2, 4.5, and 6 T, showing a spectral shift from $\Delta _{\rm SC}$ to $V_{\rm CO}$ and $\Delta ^{\prime}$ with increasing $H$. (b) Schematic of the histograms for a conventional type-II superconductor in the limit of $T \ll T_c$ and $H \ll H_{c2}$. (c) Gaussian fitting to the histograms in a reveals nearly field-independent $\Delta_{\rm SC}$ and $V_{\rm CO}$ (top), decreasing SC and increasing CO spectral weight with increasing $H$ (center), and increasing SC and CO linewidths with increasing $H$ (bottom).}
\label{Fig4}
\end{figure}

\section{Discussion}
The occurrence of two energy scales $V_{\rm CO} (> \Delta _{\rm SC})$ and $\Delta ^{\prime} (< \Delta _{\rm SC})$ inside vortices and the field-induced energy-independent wave-vectors $\textbf{Q}_{\rm PDW}$, $\textbf{Q}_{\rm CDW}$ and $\textbf{Q}_{\rm SDW}$ strongly suggest that the vortex-state quasiparticle tunneling spectra in Y-123 cannot be explained by simple Bogoliubov quasiparticle scattering interferences alone~\cite{Hanaguri09,McElroy05,ChenCT03}. On the other hand, these findings may be compared with the scenario of coexisting CO's and SC~\cite{ChenHD02,ChenHD04,Demler01,Polkovnikov02,Kivelson03,LiJX06,ChenCT07,Beyer08,Boyer07}. In particular, we find that the energy scale $V_{\rm CO}$ and the wave-vector $\textbf{Q}_{\rm CDW}$ manifested in the vortex-state spectra are consistent with the CO parameters derived from our Green's function analysis of the zero-field data. That is, we assume that the ground state of Y-123 consists of coexisting SC and CO so that the corresponding mean-field Hamiltonian is given by the following expression~\cite{ChenCT07,Beyer08}: 
\begin{eqnarray}
{\cal H}_{\rm MF} = {\cal H}_{\rm SC} + {\cal H}_{\rm CO} \qquad \qquad \qquad \qquad \qquad \qquad \qquad \nonumber\\
= \sum _{\textbf{k},\alpha} \xi _{\textbf{k}} c^{\dagger} _{\textbf{k},\alpha} c_{\textbf{k},\alpha}
- \sum _{\textbf{k}} \Delta _{\rm SC} (\textbf{k}) (c^{\dagger} _{\textbf{k},\uparrow} c^{\dagger}_{-\textbf{k},\downarrow} + c_{-\textbf{k},\downarrow} c_{\textbf{k},\uparrow}) \nonumber\\
\quad + \sum _{\textbf{k},\alpha} V_{\rm CO} (\textbf{k}) ( c^{\dagger} _{\textbf{k}+\textbf{Q},\alpha} c_{\textbf{k},\alpha} + c^{\dagger} _{\textbf{k},\alpha} c_{\textbf{k}+\textbf{Q},\alpha}). \qquad  \qquad
\label{eq:Hmf}
\end{eqnarray}
Here $\Delta _{\rm SC} (\textbf{k}) = \Delta _{\rm SC} (\cos 2 \theta _{\textbf{k}})$ for $d_{x^2-y^2}$-wave pairing, $\textbf{k}$ denotes the quasiparticle momentum, $\theta_{\textbf{k}} \equiv \tan ^{-1} (k_y / k_x)$, $\xi _{\textbf{k}}$ is the normal-state eigenenergy relative to the Fermi level, $c^{\dagger}$ and $c$ are the particle creation and annihilation operators, and $\alpha = \uparrow , \downarrow$ refers to the spin states. We incorporate realistic bandstructures into $\xi _{\textbf{k}}$ for direct comparison with experiments~\cite{ChenCT07,Beyer08}. By assuming a Fermi surface-nested CDW along $(\pi, 0)/(0, \pi)$ as the relevant CO, we diagonalize ${\cal H}_{\rm MF}$ and obtain the bare Green's function $G_0 (\textbf{k}, \omega)$. The effect of quantum fluctuations may be further included by solving the Dyson's equation for the full Green's function $G(\textbf{k},\omega)$~\cite{ChenCT07,Beyer08}. Thus, the quasiparticle DOS may be derived from $G(\textbf{k},\omega)$ as detailed in Refs.~\cite{ChenCT07,Beyer08}. For finite temperatures, we employ the temperature Green's function~\cite{ChenCT07,Beyer08}.

Following the approach outlined above and detailed in Refs.~\cite{ChenCT07,Beyer08}, we can account for the zero-field tunneling spectra shown in Fig.~1(a) by the following fitting parameters: $\Delta _{\rm SC} = (20 \pm 1)$ meV, $V_{\rm CO} = (32 \pm 1)$ meV, and $\textbf{Q} _{\rm CO}= (0.25 \pi, 0)/(0, 0.25 \pi)$ for the CDW. We note that the energy scale $V_{\rm CO} = (32 \pm 1)$ accounts for the shoulder-like features at $\omega = \pm \Delta _{\rm eff} \approx \pm 38$ meV in the tunneling spectra, where $\Delta _{\rm eff} \equiv \lbrack \Delta _{\rm SC} ^2 + V_{\rm CO} ^2 \rbrack ^{1/2}$~\cite{ChenCT07,Beyer08}. Moreover, the magnitude of $V_{\rm CO} = (32 \pm 1)$ agrees with that of the spin gap observed in neutron scattering data of optimally doped Y-123~\cite{DaiP01} and the PG energy seen inside the vortex cores, as shown in Figs.~2(e)-(g). On the other hand, the corresponding wave-vector $\textbf{Q} _{\rm CO}$ is along the Cu-O bonding direction, which remains in the vortex-state and is identified as the mode $\textbf{Q} _{\rm CDW}$ in the FT-LDOS. 

As an interesting comparison, we note that our recent vortex-state STS studies on an electron-type optimally doped cuprate $\rm La_{0.1}Sr_{0.9}CuO_2$ (La-112) also revealed PG-like features inside vortices~\cite{Teague09}, except that the PG energy in La-112 is {\it smaller} than $\Delta _{\rm SC}$. This finding may also be interpreted as a CO being revealed upon the suppression of SC. Furthermore, the smaller energy associated with the PG-like features inside vortices of La-112 is consistent with the {\it absence} of PG above $T_c$ in electron-type cuprate superconductors~\cite{Teague09}. These findings from the vortex-state quasiparticle spectra of La-112 are in contrast to those of Y-123 and the differences may be attributed to the different magnitude of $V_{\rm CO}$ relative to $\Delta _{\rm SC}$~\cite{Teague09}.   

\section{Conclusion}
Our spatially resolved scanning tunneling spectroscopic studies of Y-123 in the vortex state have revealed various novel spectral characteristics, including two energy scales ($V_{\rm CO}$ and $\Delta ^{\prime}$) other than the SC gap $\Delta _{\rm CO}$ inside vortices and three accompanying energy-independent wave-vectors $\textbf{Q}_{\rm PDW}$, $\textbf{Q}_{\rm CDW}$ and $\textbf{Q}_{\rm SDW}$. These results cannot be reconciled with theories assuming a pure SC order in the ground state~\cite{LeePA06}. Rather, they are consistent with the CO scenario and suggest important interplay between SC and various collective excitations in high-$T_c$ superconductors. 

\acknowledgments
This work was jointly supported by the Moore Foundation and the Kavli Foundation through the Kavli Nanoscience Institute at Caltech, and the NSF Grant DMR-0405088. The authors thank Dr. A. I. Rykov for growing the single crystal used in this work and Professors S. A. Kivelson and S.-C. Zhang for useful discussions. ADB acknowledges the support of Intel Graduate Fellowship.

\end{document}